# Non-contact, Real-time, Heart-rate Measurement using Image Processing with Commodity Cameras and AI Agents

**Kelly Li, Fulu Li**

**Contact Email: fulu@alum.mit.edu**

## Abstract

Heart rate measurement is one of the key requirements for real-time health monitoring, in particular for health caring of elderly people. Traditional heart rate measurement relies on contact sensing mechanisms such as some heart rate measurement devices at medical hospitals or some wearable devices with embedded sensors such as Apple Watch, etc. In this paper, we develop a system for non-contact, real-time, heart rate measurement using image processing with commodity cameras such as an embedded camera on a laptop, where we use an innovative algorithm to capture the relevant signals for the computation of heart rate in a time series in real life environments. The presented heart rate computation (HRC) process is composed with four major steps: (a) identify frames per second of the camera in use, i.e., 30 frames per second for a given camera, (b) face detection (FD) with shape predictor of 68 face landmarks using deep learning (DL) method, (c) time sliding window (TSW) algorithm to de-noise the signal by smoothing out the noise, and (d) compute heart rate based on identified signal periodicity. We test and analyze the developed prototypes against heart rate results by Apple Watch and check the difference range in multiple rounds and compute the mean of the difference for the measurement values of the heart rate of the same person at the same time. We will do further tuning and optimization of the present methods and deploy the system as a personal AI agent [6] for health monitoring as our future directions.

## 1. Introduction

Health care is one of the key pillars of our society to safeguard people's well-beings, in particular as we are going into an aging society with increasing life expectancy due to medical and public health advancements. On the other hand, fast pacing lifestyle and automation become the new norm in the era of AI (Artificial Intelligence) with LLM (Large Language Model) and deep learning[1,2,3,4,5,6]. Heart rate measurement is one of the key requirements for real-time health monitoring, in particular for elderly people.

In this paper, we develop a system for non-contact, real-time, heart rate measurement using image processing with commodity cameras, in which we build a prototype with a camera on a laptop with a simple user interface to display the computed heart rate in real time, where image frames are captured, face detection and image signals are processed. We use innovative algorithms to capture the relevant signals for the computation of heart rate in a time series in real life environments with commodity cameras such as the camera on a laptop. Our preliminary results show an error range of around 2.5% compared with Apple Watch results of the same person at the same time. we will do further optimization of the algorithms to improve the accuracy of the prototype as one of our future directions.

We discuss the related work in Section 2. We present the image processing framework in Section 3, which includes face detection, data processing, time sliding window (TSW) algorithm, and heart rate computation. We discuss application scenarios of the presented heart-rate measurement system in Section 4. We briefly touch upon the deployment of the presented system with AI agents in Section 5. Conclusions and future directions are presented in Section 6.

## 2. Related Work

In [1], H. Rahman and co-authors presented approaches for real time heart rate monitoring from facial RGB color video using webcam, where advanced signal processing algorithms like Fast Fourier Transform (FFT), Independent Component Analysis (ICA), or Principal Component Analysis (PCA) are applied to extract the Blood Volume Pulse (BVP). In [2], M. Poh and co-authors proposed non-contact, automated cardiac pulse measurements using video imaging and blind source separation, in which fast Fourier transform (FFT) are applied on the selected source signal to obtain the power spectrum. In addition, Independent Component Analysis (ICA) technique is also used to extract the related information in the computation of heart rate given video image sources. Please note that both approaches in [1] and [2] relied on the application of Fast Fourier Transform (FFT) techniques for the extraction of the related information during the process of the computation of heart rate based on image processing. In our presented approach in Section 3, we do not use Fast Fourier Transform (FFT) techniques at all, instead we present novel algorithm to extract the related information for the computation of heart rate purely in the time domain, i.e., with the time series of images in real-time video sources.

In [4], the authors discussed some image processing algorithms such as composing photomosaic images using clustering based evolutionary programming, etc. In [3], A. Vaswani and co-authors presented the seminal work of Transformer architecture that paved the way for rapid progress in Large Language Model (LLM) based generative AI. In [6], P. Steinberger presented agentic AI platform for the deployment of personal AI assistant. In [5], the authors present a comprehensive survey on multimodal Large Language Models (LLM), which may be suited for the application scenarios of the heart-rate measurement system presented in this paper.

## 3. The Image Processing Framework

In the following, we discuss the face detection (FD) method used in our image processing framework as well as the detailed workflow of data processing. We also discuss the time sliding window (TSW) algorithm to smooth out signal noice as well as the approach to compute heart rate based on signal periodicity with face images in a time series in real time video or recorded videos.

The presented non-contact, real-time, heart rate measurement (HRM) system using image processing with commodity cameras is composed with the following major components.

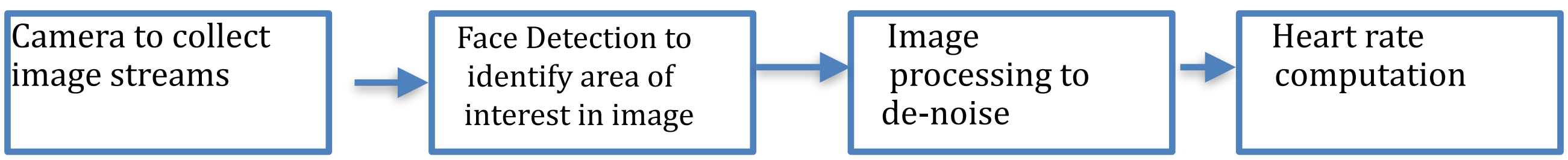

**Figure 1**: A Block Diagram of the Heart rate measurement (HRM) System

As indicated in Figure 1, the presented Heart rate measurement (HRM) system is composed with four major components: the camera to collect image streams, face detection with deep learning to identify area of interest in image, image processing to de-noise to extract relevant information for heart rate computation [1,2].

We present a novel way of developing a system with non-contact, real-time, heart rate measurement (HRM) method using image processing with commodity cameras, where we use innovative algorithm to capture the relevant signals for the computation of heart rate in a time series in real life environments such as a user sitting in front of a laptop computer with an embedded camera or a user holding an iPhone facing the iPhone camera.

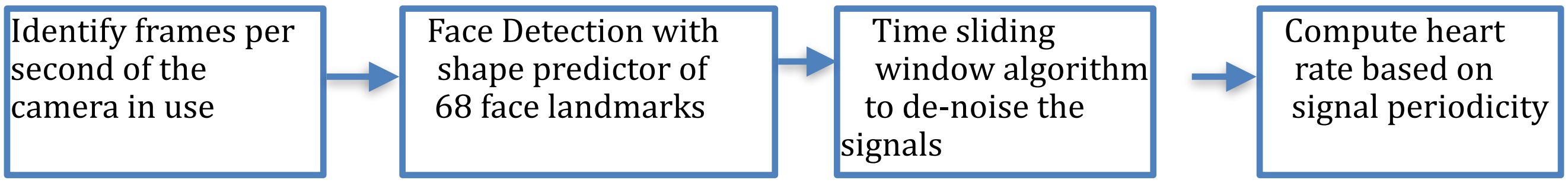

**Figure 2**: A Block Diagram of the Heart Rate Computation (HRC) Process.

As indicated in Figure 2, the heart rate computation (HRC) process is composed with four major steps: identify frames per second of the camera in use, i.e., 30 frames per second for a given camera, face detection (FD) with shaper predictor of 68 face landmarks using deep learning (DL) method, time sliding window (TSW) algorithm to de-noise the signal by smoothing out the noise, compute heart rate based on identified signal periodicity. Some related work on image processing and generative AI can be found in [3, 4, 5, 6].

In Figure 3, we show the prototype user interface (UI) that we build for non-contact, real-time, heart rate measurement (HRM) system that we developed. Basically, when we launch the program of the HRM system, it will automatically detect the face in the photo stream of the camera and the computed heart rate will appear in the image frame as well.

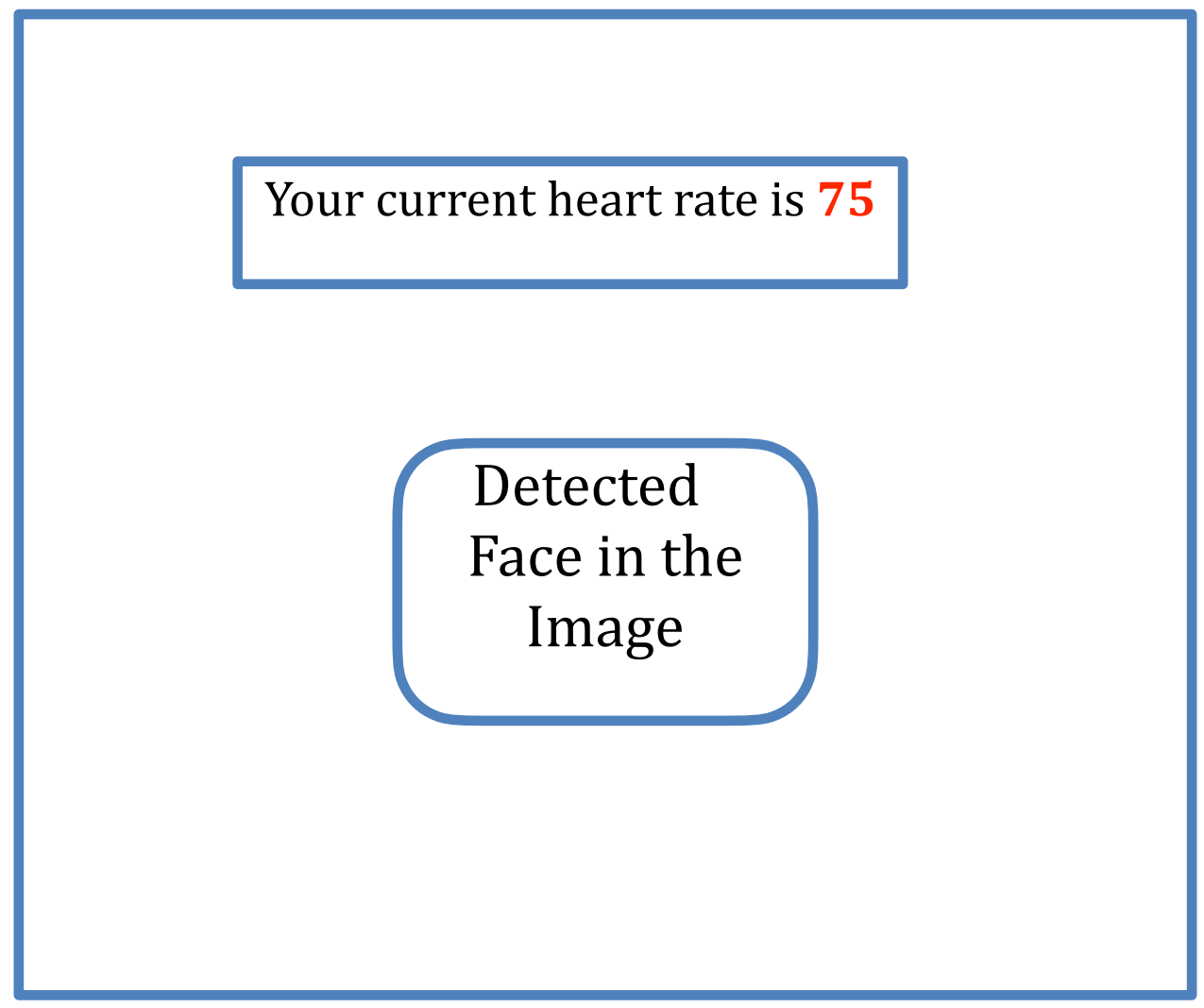

**Figure 3**: An Illustration of the Prototype User Interface (UI) of the Heart Rate Measurement (HRM) System.

We show a screenshot of the implemented prototype in Figure 4, where the measured heart rate of the user is shown in the upper left corner of the screen in real time.

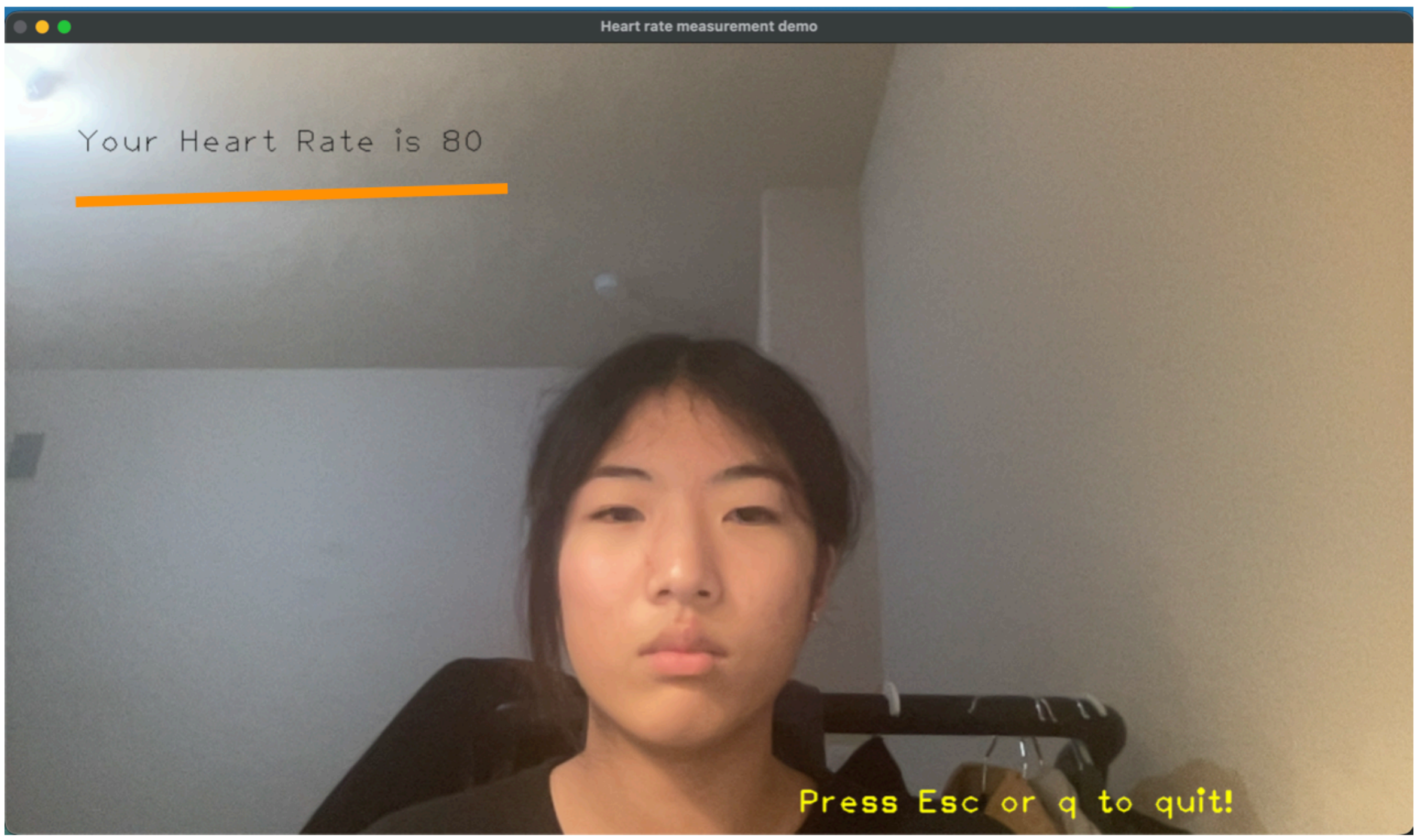

**Figure 4**: the Prototype of the Heart Rate Measurement (HRM) System.

We show experimental results of the implemented prototype compared with measurement by Apple Watch at the same time in the following table.

| Experiments | Prototype Results (heart beat rate per minute ) | Apple Watch Results (heart beat rate per minute) |
| --- | --- | --- |
| Measurement 1 | 77 | 76 |
| Measurement 2 (after small exercise) | 80 | 80 |
| Measurement 3 (after small exercise) | 82 | 80 |
| Measurement 4 (after small exercise) | 83 | 81 |
| Measurement 5 (after small exercise) | 88 | 86 |
| Measurement 6 (after small exercise) | 85 | 82 |
| Measurement 7 | 73 | 73 |
| Measurement 8 | 76 | 73 |
| Measurement 9 | 78 | 77 |
| Measurement 10 (after small exercise) | 86 | 83 |

**Figure 5:** Experimental results of the implemented prototype compared with measurement by Apple Watch of the same person at the same time. The average error range is about 2.5%. The parameters of the presented algorithm can be further tuned to improve the accuracy of the measurement of the prototype.

## 4. Application Scenarios

Application scenarios of the presented system of non-contact, real-time, heart-rate measurement using image processing with commodity cameras and agentic AI include but not limited to the following use cases: (a) emergency care to get heart-rate measurement by the patient quickly and conveniently with a cell phone or a laptop with embedded cameras; (b) virtual checkups with physicians online with video conference, where the face images of the patient can be captured during the video conference to get heart-rate measurement automatically; (c) remote monitoring of elder's health at home; (d) evaluating vehicle (bus, car, truck, etc.) driver's state to prevent traffic incidents;

## 5. Deployment with AI Agents

With the advancement of agentic AI such as the OpenClaw platform in [6] for the development and deployment of personal AI assistant, it is natural to be able to deploy the presented system of non-contact, real-time, heart-rate measurement using image processing with commodity cameras such as a laptop or an iPhone where embedded cameras are available for the deployment of the presented system. For example, the heart-rate measurement of the user can be made automatic with AI agents for the presented system based on the availability of the user's face image captured by the personal device at any time anywhere and all the related data never leave from the personal devices without the user's explicit permissions. We also need to emphasize that due to the limited computing power of the computer that we used in the experiments, we can only develop a prototype with a reasonable heart rate measurement accuracy and a reasonable delay due to image processing computations in real time for the heart rate measurement.

## 6. Conclusion and Future Directions

In this paper, we present a novel approach and develop a system for non-contact, real-time, heart rate measurement using image processing with commodity cameras such as an embedded camera on a laptop or an iPhone, where we use an innovative algorithm to capture the relevant signals for the computation of heart rate in a time series in real life environments. The presented heart rate computation (HRC) process is composed with four major steps: (a) identify frames per second of the camera in use, i.e., 30 frames per second for a given camera, (b) face detection (FD) with shape predictor of 68 face landmarks using deep learning (DL) method, (c) time sliding window (TSW) algorithm to de-noise the signal by smoothing out the noise, and (d) compute heart rate based on identified signal periodicity. We test and analyze the developed prototypes against heart rate results by Apple Watch and check the difference range in multiple rounds and compute the mean of the difference for the measurement values of the heart rate of the same person at the same time.

We will leave heart rate measurement for multiple people in the image as one of our future directions. We will also use generative AI framework to answer questions regarding the queries of the results from the system of non-contact, real-time, heart rate measurement results as one of our future research directions.